\documentclass[conference]{IEEEtran}
\usepackage{amsmath}
\usepackage{amsthm}
\usepackage{amssymb}
\usepackage{bm}
\usepackage{xspace}
\usepackage{xcolor}
\usepackage{graphicx}
\usepackage{url}
\usepackage{cite}
\usepackage{framed}
\usepackage{float}
\usepackage{rotating}

\newcommand{\executeiffilenewer}[3]{%
\ifnum\pdfstrcmp{\pdffilemoddate{#1}}%
{\pdffilemoddate{#2}}>0%
{\immediate\write18{#3}}\fi%
}

\graphicspath{{images/}}

\newcounter{algocount}

\theoremstyle{plain}

\theoremstyle{definition}

\theoremstyle{plain}

\theoremstyle{definition}

\theoremstyle{remark}

\newcommand{\vecp}{\boldsymbol{p}}

\newcommand{\vecr}{\boldsymbol{r}}

\newcommand{\vecu}{\boldsymbol{u}}
\newcommand{\vecv}{\boldsymbol{v}}
\newcommand{\vecw}{\boldsymbol{w}}

\DeclareMathOperator{\kl}{D}

\DeclareMathOperator{\entop}{\mathcal{H}}
\DeclareMathOperator{\miop}{\mathcal{I}}

\DeclareMathOperator*{\llr}{LLR}

\title{Operating LDPC Codes with Zero Shaping Gap}

\IEEEoverridecommandlockouts

\author{\IEEEauthorblockN{Georg B\"ocherer and Rudolf Mathar}
\IEEEauthorblockA{Institute for Theoretical Information
Technology\\
RWTH Aachen University, 52056 Aachen, Germany
\\ Email: \texttt{\{boecherer,mathar\}@ti.rwth-aachen.de}}
\thanks{This work has been supported by the UMIC Research Center, RWTH
Aachen University.}
}

\begin{document}
\maketitle

\begin{abstract}
Unequal transition probabilities between input and output symbols, input power constraints, or input symbols of unequal durations can lead to non-uniform capacity achieving input distributions for communication channels. Using uniform input distributions reduces the achievable rate, which is called the shaping gap. Gallager's idea for reliable communication with zero shaping gap is to do encoding, matching, and jointly decoding and dematching. In this work, a scheme is proposed that consists in matching, encoding, decoding, and dematching. Only matching is channel specific whereas coding is not. Thus off-the-shelf LDPC codes can be applied. Analytical formulas for shaping and coding gap of the proposed scheme are derived and it is shown that the shaping gap can be made zero. Numerical results show that the proposed scheme allows to operate off-the-shelf LDPC codes with zero shaping gap and a coding gap that is unchanged compared to uniform transmission.
\end{abstract}

\section{Introduction}

The ultimate rate for reliable communication over a noisy channel is given by the maximum mutual information between input and output. A channel is called non-uniform if the input distribution that achieves this maximum is non-uniform. The non-uniformity can result from different factors, examples are channels that are not symmetric \cite[Theorem~8.2.1]{Cover2006}, average power constraints on input symbols \cite[Sec.~IV]{Blahut1972}, input symbols of unequal duration \cite{Jimbo1979}, and multi-user channels with cross-talk \cite{Ratzer2003a}. Conventional communication systems restrict the channel input to be uniformly distributed. The resulting penalty in terms of achievable rates is called the shaping gap.

One approach to close the shaping gap goes back to Gallager \cite[p. 208]{Gallager1968} and is nicely explained by McEliece in \cite[Sec.~5]{McEliece2001}. The idea is to first encode the data and then to use a fixed-to-fixed length mapper from the codewords to the channel symbols. The mapping is chosen such that the capacity achieving probability mass function (pmf) is approximated. To achieve this, the mapping is many-to-one. For iterative decoding, a probabilistic demapper needs to be incorporated into the factor graph of the decoder \cite[Fig.~8]{McEliece2001}. This idea was extended to non-binary low-density parity-check (LDPC) codes by Ratzer and MacKay in \cite{Ratzer2003} and by Bennatan and Burshtein in \cite{Bennatan2004}.

Franceschini \emph{et al.} conjectured in \cite{Franceschini2006} that LDPC codes have universal properties, which basically means the following: \emph{if an LDPC code is designed for some channel A, it can be used for a different channel B, as long as the mutual information between input and output are the same both for channel A and channel B}. See \cite{Sason2011} and references therein for analytical support of this conjecture. 

In this work, we use the conjecture of Franceschini \emph{et al.} as a design paradigm. That is, instead of designing LDPC codes for specific non-uniform channels, \emph{we show how existent LDPC encoders/decoders that were originally designed for uniform channels can be operated with zero shaping gap on non-uniform channels}.

We propose the following transmission scheme. First, the binary equiprobable data stream is matched to the channel by a prefix-free code \cite{Bocherer2011}. Then, the matched bits are encoded by a systematic code. The check bits are iteratively matched to the channel and encoded such that all transmitted bits are matched to the channel. At the receiver side, the packets are first decoded and then dematched. The central property of our scheme is that shaping (matcher) and error correction (encoder/decoder) can be designed independently, in the sense that the matcher only influences the \emph{parameters} passed to the encoder (i.e., data symbols) and decoder (i.e., log-likelihood ratios (LLR) and priors). This is a fundamental difference to Gallager's scheme, where decoding and dematching has to be performed jointly. For binary channels with unequal symbol durations, we provide analytical formulas for the shaping gain and the coding gain of our scheme. We show that capacity achieving matching codes can efficiently be found using Geometric Huffman Coding \cite{Bocherer2011}. Our results can easily be extended to arbitrary discrete input memoryless channels. Finally, we apply our scheme to a binary symmetric channel (BSC) with symbol durations $w_0=1$ and $w_1=5$ and varying error probability $\epsilon$. We use off-the-shelf LDPC codes, namely the codes used in DVB-S2 as implemented in the Matlab Communications Toolbox. For uniform transmission, we observe a shaping gap of $20\%$ while for matched transmission, the shaping gap is virtually zero. For both schemes, the coding gap is almost equal, which is in perfect accordance with Francheschini's conjecture.

The remainder of this paper is organized as follows. In Section~\ref{sec:system}, we define the binary channel with unequal symbol durations. In Section~\ref{sec:bootstrap}, we develop the proposed scheme. Formulas for the shaping and coding gain are derived in Section~\ref{sec:gain}. Capacity achieving matchers are defined in Section~\ref{sec:matching}. Finally, we provide numerical results in Section~\ref{sec:numeric}.

\section{Problem Statement}
\label{sec:system}
\begin{figure*}
\centering
\def\svgwidth{0.75\textwidth}
\footnotesize
\executeiffilenewer{images/bootstrap_system.svg}{images/bootstrap_system.pdf}%
{inkscape -z -D --file=images/bootstrap_system.svg %
--export-pdf=images/bootstrap_system.pdf --export-latex}%
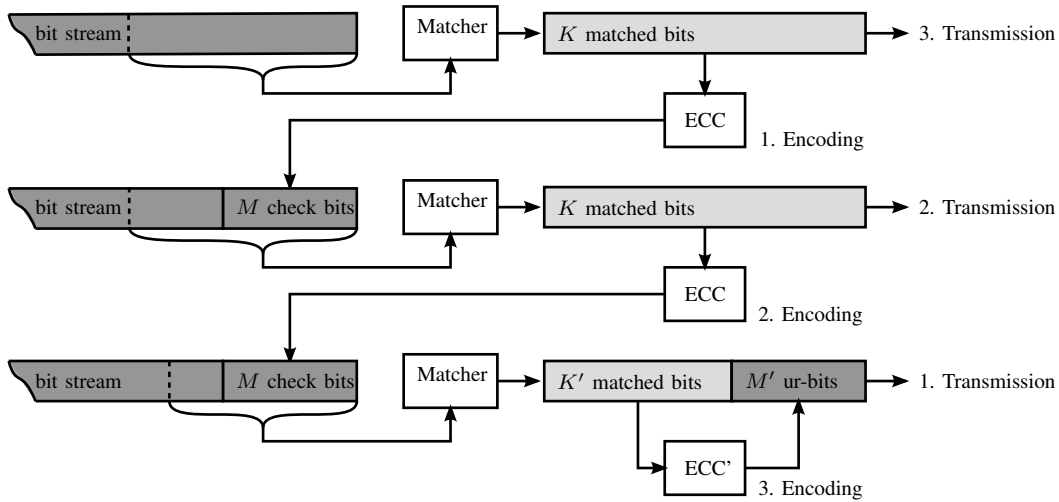%

\caption{Bootstrap scheme. The acronym ECC stands for error correction coding.}
\label{fig:bootstrap}
\end{figure*}
A binary channel (BC) is specified by a matrix of transition probabilities $(h_{ji})$. An input PMF $\vecp$ relates to its corresponding output PMF $\vecr$ as
\begin{align}
\vecr=
\left(
\begin{array}{c}
r_0\\
r_1
\end{array}
\right)
=
\left(
\begin{array}{ccc}
h_{00}&h_{01}\\
h_{10}&h_{11}
\end{array}
\right)
\left(
\begin{array}{c}
p_0\\
p_1
\end{array}
\right)
.
\end{align}
The mutual information between input and output is according to \text{Gallager \cite[Eq. (8.73)]{Gallager2008}} given by
\begin{align}
\miop(\vecp)&=\sum_i p_i\sum_j h_{ji}\log\frac{h_{ji}}{r_j}.
\end{align}
We consider BCs where the input symbols $0$ and $1$ have unequal durations $\vecw=(w_0,\,w_1)^T$. According to \cite[Theorem~2]{Verdu1990}, the capacity of such a BC is given by
\begin{align}
\label{prob:dmcDurations}
\mathsf{C}=\max_{\vecp}\frac{\miop(\vecp)}{\vecw^T\vecp}.
\end{align}
Jimbo and Kunisawa \cite{Jimbo1979} proposed a variation of the Blahut-Arimoto algorithm \cite{Blahut1972,Arimoto1972} to efficiently find the capacity achieving distribution $\vecp^*$  that maximizes \eqref{prob:dmcDurations}. We denote the uniform pmf by $\vecu=(0.5,\,0.5)^T$. The shaping gain that results from using $\vecu$ instead of $\vecp^*$ is given by
\begin{align}
\frac{\miop(\vecu)}{\vecw^T\vecu}\frac{1}{\mathsf{C}}.
\end{align}
In the case of equal symbol durations $w_0=w_1$, this gain is at least $0.942$ \cite[p. 295]{Richardson2008} and there are little reasons not to use $\vecu$. However, for $w_0\neq w_1$, the gain can be arbitrary close to zero. This motivates us to seek for a transmission scheme that closes the shaping gap for arbitrary symbol durations $\vecw$ while allowing to use existent LDPC encoders/decoders.

\section{Bootstrap Scheme}
\label{sec:bootstrap}
\subsection{Prefix-Free Matcher Codes}
The digital interface between source and channel coding is a stream of independent, identically distributed (iid) equiprobable bits. By parsing the stream by a full prefix-free code, a dyadic pmf can be generated \cite{Bocherer2011}. For example, consider the set of binary input symbol blocks of length $2$, namely $\{00,01,10,11\}$. Then the mapping
\begin{align}
\begin{array}{rr}
1 \mapsto 00&
01 \mapsto 01\\
001 \mapsto 10&
000 \mapsto 11
\end{array}\label{eq:matcher}
\end{align}
generates the pmf $(2^{-1},2^{-2},2^{-3},2^{-3})$ over the set $\{00,01,10,11\}$ when the stream of iid equiprobable bits is parsed by the prefix-free code $\{1,01,001,000\}$. We call a device that implements this procedure a \emph{matcher}. We will see in Section~\ref{sec:matching} how capacity achieving matchers can be found efficiently. When matchers are used for noisy channels, a severe problem occurs: one single bit error can lead to a complete loss of a block, since the binary input and output streams are out of sync, e.g., suppose the data bits $101001$ were mapped by the matcher \eqref{eq:matcher} to the block $000110$ and then transmitted over the channel, and suppose $010110$ was detected at the channel output. Then,
\begin{align}
&101001\mapsto 000110\\
010110\mapsto &0101001
\end{align}
i.e., matcher input and dematcher output are of different length and aligning the first two bits leads to $5$ bit errors in the overlapping strings. Error-correction based on matcher input and dematcher output needs the capability of correcting insertion and deletion errors, which is difficult \cite{Mitzenmacher2008}. 
\subsection{Reverse Concatenation}
The above problem can be solved by interchanging the order of error-correction coding and matching. This approach is used for constrained systems by first applying constrained coding and then error correction coding, see \cite{Blaum2007} and references therein. Suppose we have applied the matcher and apply error correction coding to the matcher output. In LDPC codes, sums of bits modulo $2$ are transmitted. As the number of summands grows, the resulting bit is uniformly distributed independent of the pmf of the summands, i.e.,
\begin{align}
p = i_1\oplus i_2 \oplus \dotsb\oplus i_m \sim U\{0,1\}.
\end{align}
Thus, even if the bits $i_1,\dotsc,i_m$ were matched to the channel, $p$ is not. A first step to circumvent this problem is to use systematic codes. Here, the data bits are transmitted unchanged and the check bits are appended. Thus, the data bits can be shaped and the check bits cannot. This already gives an improvement compared to uniform transmission. This approach was suggested under the name \emph{sparse-dense codes} by Ratzer in \cite{Ratzer2003}. However, this approach does not close the shaping gap, e.g., for 1/2-rate codes, only half of the transmitted bits are matched to the channel.
\subsection{Bootstrapping the Check Bits}
We now introduce the \emph{bootstrap scheme}, which allows to match the check bits to the channel. Consider a systematic binary block code that generates $M$ check bits per $K$ data bits, i.e., a rate $K/(K+M)$-code. At the transmitter side, $B$ blocks are sequentially encoded and then transmitted in reverse order. For the first block, the matcher generates $K$ matched bits from the equiprobable input bit stream. From these $K$ matched bits, the encoder calculates $M$ check bits. These check bits are approximately iid and equiprobable. They are fed back and appended to the bit stream. In the next round, the matcher again generates $K$ matched bits from the bit stream. Since the $M$ check bits from the first round were appended to the bit stream, they are now matched to the channel as part of the $K$ matched bits of the second round. This procedure continuous until the last but second round. In the last round, a $(M',K')$ sparse-dense code is applied. The matcher maps the $M$ check bits from round $B-1$ plus data bits to $K'$ matched bits. The encoder calculates $M'$ check bits from the $K'$ matched bits and appends them (unmatched) to the $K'$ matched bits. These $M'$ check bits of the last round are the \emph{ur-bits} from which the decoder will start to bootstrap all $B$ blocks. All $B$ blocks are transmitted in reverse order such that the decoder can immediately start decoding. See Fig.~\ref{fig:bootstrap} for an illustration.

\begin{figure}
\centering
\def\svgwidth{1.0\columnwidth}
\footnotesize
\executeiffilenewer{images/building_block.svg}{images/building_block.pdf}%
{inkscape -z -D --file=images/building_block.svg %
--export-pdf=images/building_block.pdf --export-latex}%
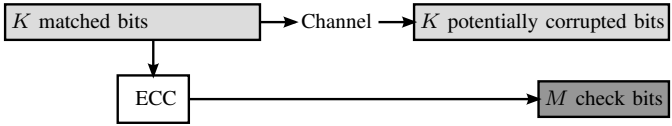%

\caption{Building block of the bootstrap scheme.}
\label{fig:building}
\end{figure}
\section{Shaping and Coding Gain}
\label{sec:gain}
The number $M'$ of ur-bits is independent of the number of packets $B$. Thus, as $B$ grows, the fraction of matched bits approaches $1$. Therefore, we will focus in the rest of this work on the building block of the bootstrap scheme, which consists of the transmission of $K$ matched bits over a channel and decoding conditioned on $M$ perfectly known check bits. See Fig.~\ref{fig:building} for an illustration. Note that the $M$ perfectly known check bits are not for free: the $M$ check bits of the next packet to be received are embedded in matched form in the $K$ matched bits. We will take this into account when calculating the effective transmission rate.

\subsection{Shaping Gain}
Assume the matcher uses a variable-to-$k$ length code, i.e., blocks of $k$ matched bits are jointly generated from the unmachted bits. Denote by $\vecp$ the resulting block pmf. Denote by $\vecv$ the lengths of the blocks, e.g., if block $i$ consists of $\ell$ zeros and $k-\ell$ ones, the length is $v_i=\ell w_0+(k-\ell)w_1$. The mutual information rate is directly given by
\begin{align}
\mathsf{I}_\mathrm{bs}(\vecp)=\frac{\miop(\vecp)}{\vecv^T\vecp}
\end{align}
and the shaping gain is given by
\begin{align}
\frac{\mathsf{I}_\mathrm{bs}(\vecp)}{\mathsf{C}}.
\end{align}
\subsection{Coding Gain}
We now calculate the effective transmission rate of the bootstrap scheme, i.e., how many information bits per block length are in the average transmitted over the channel. Denote by $\entop(\vecp)$ the entropy in bits of $\vecp$. The average information per matched bit is $\entop(\vecp)/k$. Thus, in the average,
\begin{align}
\frac{\sharp\{\text{unmachted bits}\}}{\sharp\{\text{matched bits}\}}=\frac{1}{\frac{\entop(\vecp)}{k}}\triangleq m.
\end{align}
The unmachted bits consist of the information bits and the $M$ check bits of the previous block. Thus
\begin{align}
(\sharp\{\text{information bits}\}&+M)m=\sharp\{\text{matched bits}\}=K\\
\Rightarrow&\sharp\{\text{information bits}\}=\frac{K}{m}-M.
\end{align}
The average length of a matched bit is $\vecv^T\vecp/k$. Denote by $c=K/(K+M)$ the code rate of the applied code. Altogether, the effective transmission rate can now be written as
\begin{align}
\frac{\sharp\{\text{information bits}\}}{K\frac{\vecv^T\vecp}{k}}&=\frac{\frac{K}{m}-M}{K\frac{\vecv^T\vecp}{k}}\\
&=\frac{\frac{\entop(\vecp)}{k}-\frac{M+K-K}{K}}{\frac{\vecv^T\vecp}{k}}\\
&=\frac{\frac{\entop(\vecp)}{k}+1-\frac{1}{c}}{\frac{\vecv^T\vecp}{k}}\\
&\triangleq\mathsf{R}_\mathrm{bs}(\vecp,c).
\end{align}
We define the \emph{coding gain} by
\begin{align}
\frac{\mathsf{R}_\mathrm{bs}(\vecp,c)}{\mathsf{I}_\mathrm{bs}(\vecp)}=\frac{\frac{\entop(\vecp)}{k}+1-\frac{1}{c}}{\frac{\miop(\vecp)}{k}}.
\end{align}

\section{Capacity Achieving Matcher}
\label{sec:matching}
We now show that the shaping gap can indeed be made as small as desired by approximating the capacity achieving pmf $\vecp^*$ with the matching code found via Geometric Huffman Coding (\textsc{Ghc}) \cite{Bocherer2011}. We show this by first calculating the penalty that results from using an input pmf different from $\vecp^*$, and second, by bounding this penalty.
\subsection{Using the Wrong PMF}
The capacity achieving pmf $\vecp^*$ is efficiently found by the Algorithm of Jimbo and Kunisawa \cite{Jimbo1979}. By \cite[Lemma~2]{Jimbo1979},
\begin{align}
\sum_j h_{ji}\log\frac{h_{ji}}{p^*_j} &\leq  \mathsf{C} w_i,\qquad\text{with equality if } p_i^* > 0.\label{eq:kktDmcProperty}
\end{align}
Denote by $\vecp$ some pmf with the only restriction that
\begin{align}
p_i=0,\qquad\text{whenever }p_i^*=0.\label{eq:PmfCondition}
\end{align}
Now,
\begin{align}
\miop(\vecp)&=\sum_i p_i\sum_j h_{ji}\log\frac{h_{ji}}{r_j}\\
&=\sum_i p_i\sum_j h_{ji}\log\frac{h_{ji}r_j^*}{r_jr_j^*}\\
&=\sum_i p_i\sum_j h_{ji}\log\frac{h_{ji}}{r_j^*}+\sum_i p_i\sum_j h_{ji}\log\frac{r_j^*}{r_j}\\
&=\mathsf{C}\vecw^T\vecp-\kl(\vecr\Vert\vecr^*)
\end{align}
where equality in the last line follows from \eqref{eq:PmfCondition} and \eqref{eq:kktDmcProperty}. Dividing by $\vecw^T\vecp$ yields
\begin{align}
\frac{\miop(\vecp)}{\vecw^T\vecp}=\mathsf{C}-\frac{\kl(\vecr\Vert\vecr^*)}{\vecw^T\vecp}.\label{eq:wrongPmf}
\end{align}
Thus, using pmf $\vecp$ instead of the capacity achieving pmf $\vecp^*$ results in a penalty of $\frac{\kl(\vecr\Vert\vecr^*)}{\vecw^T\vecp}$.
\subsection{Capacity Achieving Prefix-Free Matcher}
By \cite[Eq. (4.45)]{Cover2006}, the KL-distance between the output pmfs $\vecr$ and $\vecr^*$ is upper bounded by the KL-distance between the corresponding input pmfs, i.e., $\kl(\vecr\Vert\vecr^*)\leq\kl(\vecp\Vert\vecp^*)$. Thus, the penalty is upper bounded by
\begin{align}
\frac{\kl(\vecp\Vert\vecp^*)}{\vecw^T\vecp}.\label{eq:penaltyBound}
\end{align}
By jointly approximating the capacity achieving pmf of $k$ consecutive input symbols via \textsc{Ghc}, by \cite[Prop.~2]{Bocherer2011}, the penalty bound \eqref{eq:penaltyBound} goes to zero as block-length $k$ goes to infinity. Consequently, the left-hand side of \eqref{eq:wrongPmf} approaches capacity.

\section{Application to a BSC with $\vecw=(1,\,5)^T$}
\label{sec:numeric}
We now consider a binary symmetric channel (BSC) where the input symbols have the weights $w(0)=1$ and $w(1)=5$. We vary the bit error probability $\epsilon$ between $0.005$ and $0.057$. As LDPC codes, we use the DVB-S2 codes with rates $8/9$, $5/6$, $4/5$, and $3/4$. The DVB-S2 codes are systematic and amenable for simulation, since an implementation is readily available in the Communications System Toolbox of Matlab. Because the simulation is very time-intensive, we evaluate the performance around block-error rates $p_b$ of $10^{-2}$. For each value of $\epsilon$, capacity achieving pmf $\vecp^*$ and capacity $\mathsf{C}$ are calculated using the Jimbo-Kunisawa Algorithm \cite{Jimbo1979}. Both for uniform and matched transmission, we use exactly the same parity check matrices, encoders, and decoders, namely the unchanged implementation in Matlab. The differences between uniform and matched transmission are detailed next.
\subsection{Uniform Transmission}
\subsubsection{Matching} $K$ iid equiprobable data bits are generated. No matching is performed.
\subsubsection{Transmission} Both the $K$ data bits and the $M$ check bits are transmitted over a BSC with parameter $\epsilon$.
\subsubsection{Decoder parameters}
Since we have uniform priors, we pass to the decoder for each of the $K=K+M$ received bits the LLRs
\begin{align*}
\llr(0)=\ln\frac{(1-\epsilon)}{\epsilon}, \text{ if received bit }y=0\\
\llr(1)=\ln\frac{\epsilon}{(1-\epsilon)}, \text{ if received bit }y=1.
\end{align*}
\subsubsection{Evaluation}
For each $\epsilon$, the shaping gain is according to (4) calculated as
\begin{align}
\frac{\miop(\vecu)}{\vecw^T\vecu}\frac{1}{\mathsf{C}}
\end{align}
where $\vecu$ is the uniform pmf. For each coding rate $c$ and channel parameter $\epsilon$, the coding gain calculates as
\begin{align}
\frac{c}{\miop(\vecu)}.
\end{align}
\subsection{Matched Transmission}
\begin{figure}
\centering
\footnotesize
\begin{tabular}{rrrr}
$00\mapsto 0000$&$010\mapsto 0001$&$011\mapsto 0010$&$10111\mapsto 0011$\\
$100\mapsto 0100$&$10110\mapsto 0101$&$10100\mapsto 0110$&\!\!$1010111\mapsto 0111$\\
$110\mapsto 1000$&$11101\mapsto 1001$&$11100\mapsto 1010$&$101011\mapsto 1011$\\
\!\!\!\!$11110\mapsto 1100$&\!\!$111110\mapsto 1101$&\!\!$111111\mapsto 1110$&\!\!$1010100\mapsto 1111$
\end{tabular}
\caption{Matching code found by \textsc{Ghc}. The same matching code was found for the whole considered range of the error probability $\epsilon$.}
\label{fig:matcher}
\vspace{-0.5cm}
\end{figure}
\subsubsection{Matching}
For the capacity achieving joint pmf $\vecp^{*4}$ of $4$ subsequent bits, the dyadic approximation $\vecp$ is calculated using the implementation \cite{website:ghc} of $\textsc{Ghc}$. The prefix-free matching code induced by $\vecp$ is displayed in Fig.~\ref{fig:matcher}. For the whole considered range of $\epsilon$, the optimal dyadic approximation remains unchanged. A sufficiently long data stream of iid equiprobable bits is generated. By parsing the stream by the matcher code, a sequence of $K/4$ blocks consisting each of $4$ bits is generated. These blocks are iid according to $\vecp$.
\subsubsection{Transmission} We transmit the $K$ matched data bits over a BSC with parameter $\epsilon$. The $M$ check bits remain unchanged.
\subsubsection{Decoder}
Since we have non-uniform priors $\vecp^*=(\pi_0,\,\pi_1)^T$, we pass to the decoder for the $K$ received bits
\begin{align*}
\llr(0)+\ln\frac{\pi_0}{\pi_1}=\ln\frac{(1-\epsilon)\pi_0}{\epsilon\pi_1}, \text{ if received bit }y=0\\
\llr(1)+\ln\frac{\pi_0}{\pi_1}=\ln\frac{\epsilon\pi_0}{(1-\epsilon)\pi_1}, \text{ if received bit }y=1
\end{align*}
and for the $M$ check bits, we pass
\begin{align*}
\infty, &\text{ if check bit }b=0\\
-\infty, &\text{ if check bit }b=1.
\end{align*}
\subsubsection{Evaluation}
For each channel parameter $\epsilon$, the shaping gain is according to (10) calculated as
\begin{align}
\frac{\miop(\vecp)}{\vecv^T\vecp}\frac{1}{\mathsf{C}}
\end{align}
where $\vecv$ denotes the weights of the symbol blocks. For each coding rate $c$ and channel parameter $\epsilon$, the coding gain is according to (18) calculated as
\begin{align}
\frac{\frac{\entop(\vecp)}{4}+1-\frac{1}{c}}{\frac{\miop(\vecp)}{4}}.
\end{align}

\subsection{Discussion}
\begin{figure}
\centering
\def\svgwidth{0.98\columnwidth}
\footnotesize
\executeiffilenewer{images/uniform_gain.svg}{images/uniform_gain.pdf}%
{inkscape -z -D --file=images/uniform_gain.svg %
--export-pdf=images/uniform_gain.pdf --export-latex}%
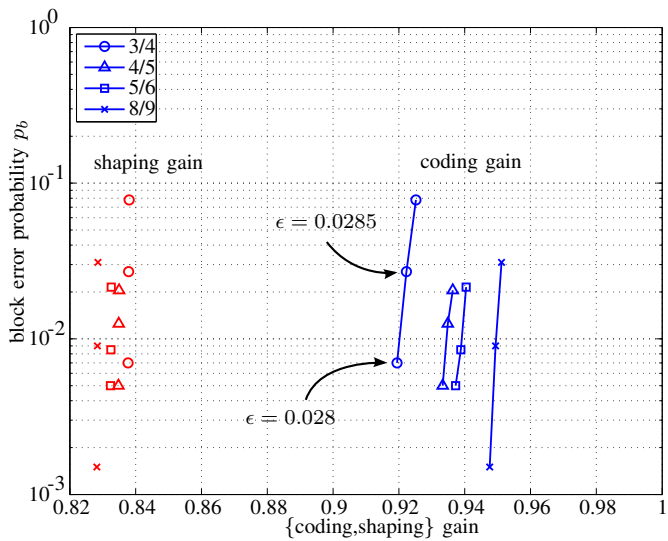%

\caption{Uniform transmission.}
\label{fig:uniformResults}
\end{figure}

\begin{figure}
\def\svgwidth{0.98\columnwidth}
\footnotesize
\executeiffilenewer{images/bootstrap_gain.svg}{images/bootstrap_gain.pdf}%
{inkscape -z -D --file=images/bootstrap_gain.svg %
--export-pdf=images/bootstrap_gain.pdf --export-latex}%
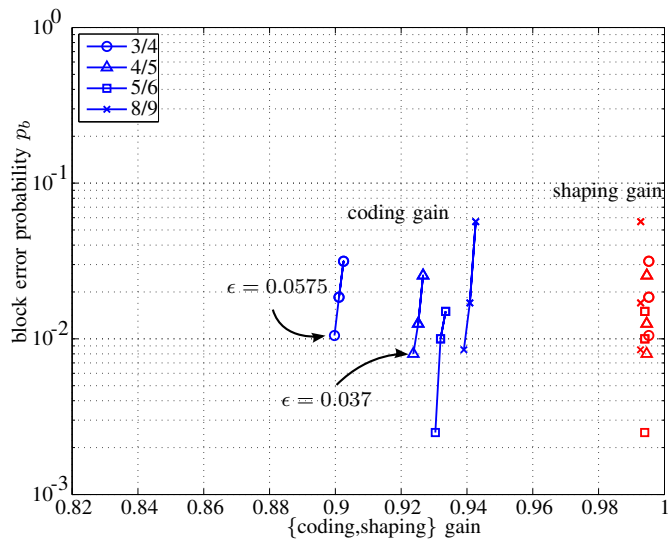%

\caption{Matched transmission.}
\label{fig:bootstrapResults}
\end{figure}

The simulation results are displayed in Fig.~\ref{fig:uniformResults} and Fig.~\ref{fig:bootstrapResults}. Shaping and coding gain versus block error probability $p_b$ are displayed for $3/4$, $4/5$, $5/6$, and $8/9$ DVB-S2 codes. The different operating points were generated by varying the channel error probability $\epsilon$. For uniform transmission, the shaping gain is around $0.83$. For matched transmission, the shaping gain is larger than $0.99$. The coding gain is for uniform transmission between $0.92$ and $0.95$. For matched transmission, it is between $0.9$ and $0.94$. For example, for the rate $3/4$ code, the coding gain for uniform transmission is $0.92$ and for matched transmission, it is $0.9$, thus slightly worse. However, for uniform transmission, the $3/4$ code works at $p_b=10^{-2}$ for some $\epsilon$ between $0.028$ and $0.0285$ , while for matched transmission, the $3/4$ code works at $p_b=10^{-2}$ for $\epsilon=0.0575$ , which is a significantly higher channel error probability. If we compare the achieved coding gains of uniform and matched transmissions at similar $(p_b,\epsilon)$ pairs, we observe a higher coding gain for matched transmission than for uniform transmission. For example, we have a coding gain of $0.921$ at $(p_b=0.03,\epsilon=0.0285)$ for uniform transmission and a coding gain of more than $0.921$ at $(p_b=0.008,\epsilon=0.037)$. This corresponds to a better coding gain at a lower block error rate for a higher channel error probability.

\bibliographystyle{IEEEtran}
\normalsize
\bibliography{IEEEabrv,confs-jrnls,Dnc}

\end{document}